\documentclass[a4paper, 11pt]{article}
\usepackage{cite}
\usepackage{pos}
\usepackage{physics}
\usepackage{url}
\usepackage[frozencache,cachedir=minted-files]{minted}
\usepackage{wrapfig}
\usepackage{subcaption}
\usepackage{graphicx}

\usepackage{color}
\usepackage{tikz}
\usetikzlibrary{decorations.pathmorphing,decorations.markings}
\usetikzlibrary{decorations.pathreplacing}
\usetikzlibrary{positioning, calc, shapes}
\usetikzlibrary{shapes.geometric, arrows, calc}
\tikzstyle{tool} = [rectangle, rounded corners, minimum width=1cm, text
					centered, draw=black, fill=red!30]
\tikzstyle{func} = [rectangle, dashed, minimum width=1cm, minimum height=0.6cm,
					text centered, draw=black, fill=green!30]
\tikzstyle{class} = [rectangle, rounded corners, minimum height=0.6cm, minimum
					 width=1cm,text centered, draw=black, fill=blue!30]
\tikzstyle{block} = [rectangle, dashed, minimum width=1cm, minimum height=0.6cm,
					 text centered, draw=black, fill=white]

\tikzstyle{arrow} = [thick,->,>=stealth]
\tikzstyle{line} = [-,>=stealth]

\usepackage[compat=1.1.0]{tikz-feynman}
\usepackage{contour}

\definecolor{darkgreen}{rgb}{0.0, 0.5, 0.13}
\definecolor{darkred}{rgb}{0.55, 0.0, 0.0}
\title{Extending \texttt{MadFlow}: device-specific optimization}
\ShortTitle{\texttt{MadFlow}}

\author[a,b,c]{Stefano Carrazza}
\author*[a,b]{Juan M. Cruz-Martinez}
\author[a,d]{Gabriele Palazzo}

\affiliation[a]{TIF Lab, Dipartimento di Fisica, Universit\`a degli Studi di
                Milano and INFN Sezione di Milano \\
                Via Celoria 16, 20133, Milano, Italy}
\affiliation[b]{CERN, Theoretical Physics Department, CH-1211 Geneva 23, Switzerland}
\affiliation[c]{Quantum Research Centre, Technology Innovation Institute, Abu Dhabi, UAE.}
\affiliation[d]{Medical Physics, San Raffaele Scientific Institute, Milano, Italy}

\emailAdd{stefano.carrazza@mi.infn.it}
\emailAdd{juan.cruz.martinez@cern.ch}
\emailAdd{palazzo.gabriele@hsr.it}

\abstract{In this proceedings we demonstrate some advantages of a top-bottom approach in the development of hardware-accelerated code.
We start with an autogenerated hardware-agnostic Monte Carlo generator, which is parallelized in the event axis. This allow us to take advantage of the parallelizable nature of Monte Carlo integrals even if we don't have control of the hardware in which the computation will run (i.e., an external cluster).
The generic nature of such an implementation can introduce spurious bottlenecks or overheads.
Fortunately, said bottlenecks are usually restricted to a subset of operations and not to the whole vectorized program. By identifying the more critical parts of the calculation one can get very efficient code and at the same time minimize the amount of hardware-specific code that needs to be written. We show benchmarks demonstrating how simply reducing the memory footprint of the calculation can increase the performance of a $2 \to 4$ process.}

\FullConference{%
  41st International Conference on High Energy physics - ICHEP2022\\
  6-13 July, 2022\\
  Bologna, Italy
}

\note{\textit{Preprint numbers:}}
\begin{document}

\maketitle

\section{Introduction and motivation}
\label{sec:introduction}
Research in high energy physics has historically been tied very closely to the development of high performance computing.
This is still true today; obtaining physical predictions for some of the most accurate and precise calculations can take many CPU-years~\cite{Britzger:2022lbf} due to the very complicate nature of the most interesting quantities. 
One example of a computationally expensive application in particle physics is event generators~\cite{Campbell:2022qmc} which often make use of Monte Carlo methods.

In order to reduce the computing cost of a Monte Carlo integration several strategies can be followed, the most obvious of which is a simple (but effective) optimization of the existing code and algorithms, making the computation faster but otherwise exactly the same.
Perhaps the most interesting and promising venue instead is to develop new algorithms, although this is also the most challenging path.
Despite many attempts in recent years~\cite{DBLP:disneypaper,Bothmann:2020ywa,Gao:2020zvv,Gao:2020vdv}, the Vegas algorithm~\cite{Lepage:1977sw,Lepage:1980dq} remains unbeaten in terms of use, specially when taking into account the many-dimensional space over which these integrals are calculated.

Another possibility, which can be seen as an intermediate path between the previous approaches, is to modify current algorithms to target different types of hardware, potentially enabling enormous speed-ups.
This is something that has become possible only recently thanks to General Purpose GPU computing and the development of new hardware devices such as Tensor Processing Units (TPUs).

Based on familiar tools (tensors) and on an intuitive programming language (\texttt{python}), the techniques described in Ref.~\cite{Carrazza:2020rdn} allow physicists to write programs that can run on both CPUs and hardware accelerators, obtaining enormous speed-ups without having to deal with the peculiarities of any particular device or architecture.
Furthermore, the code can be easily interfaced to other programs and libraries often used in HEP~\cite{Carrazza:2020pkv}, allowing for great flexibility.

In Ref.~\cite{Carrazza:2020rdn} we introduced \texttt{VegasFlow}, a framework to facilitate the transition from CPU-based computations to GPU for developers familiarized with Monte Carlo integration.
\texttt{VegasFlow} is a reimplementation of the original Vegas algorithm in a vectorized (or rather, {\it tensorized}) manner such that the events necessary for the calculation are computed in parallel.
Since all events are parallel from the onset (instead of the parallelization of an existing sequential code) the usage of hardware accelerators, such as GPUs, becomes natural.
In addition, the code is written using TensorFlow~\cite{tensorflow2015:whitepaper}, a machine learning framework which already includes many primitives for everyday mathematical operations with kernels able to run in CPUs and GPUs of different manufacturers.

In Ref.~\cite{Carrazza:2020qwu} the aforementioned techniques were used to write \texttt{PDFFlow}, an adaptation of the well-known {\tt LHAPDF} library~\cite{Buckley:2014ana} which exploits the massive capacity of modern GPUs for PDF interpolation.
This is a crucial part of a simulation of proton collisions.
As an example, in the same publication, a prototype of a Next-to-Leading Order calculation running entirely in a GPU was presented.

Finally, in Ref.~\cite{Carrazza:2021gpx} we presented \texttt{MadFlow}.
\texttt{MadFlow} includes a Madgraph plugin
which takes advantage of {\tt MG5\_aMC}'s~\cite{Alwall:2014hca} great flexibility and
extensibility to autogenerate matrix-elements completely {\it tensorized} in the
event dimension, and thus ripe to be offloaded to a hardware accelerator.

In each of the previous steps the flexibility and user-friendliness of the code have been primed with respect to performance.
Nevertheless, as it has been demonstrated in several benchmarks in the cited publications, the hardware-accelerated calculation has often been proven to be more efficient than the CPU-based one (at equal cost and even without taking into consideration more specific optimizations).
However, as it is usually the case, the price to pay for convenience and comfort is a high-level code that cannot truly take full advantage of the hardware as much as a low-level targeted implementation could.
High-level code often introduces spurious bottlenecks which are not fundamental to the underlying algorithms.

In this proceedings we demonstrate how surgically eliminating some of these bottlenecks can further unlock the potential of hardware accelerators while keeping most of the convenience of having a generic codebase.
This is akin to a top-bottom approach in which we optimize at low-level only the parts of the computation where it is actually needed.
By reducing the scope of what needs to be implemented at low-level one can more easily leverage the features of the available hardware.
In exchange, we lose the flexibility as different hardware architectures allow for different kinds of optimizations or programming languages.

\section{Technical details}
\label{sec:techimp}
\begin{figure}
    \includegraphics[width=0.49\textwidth]{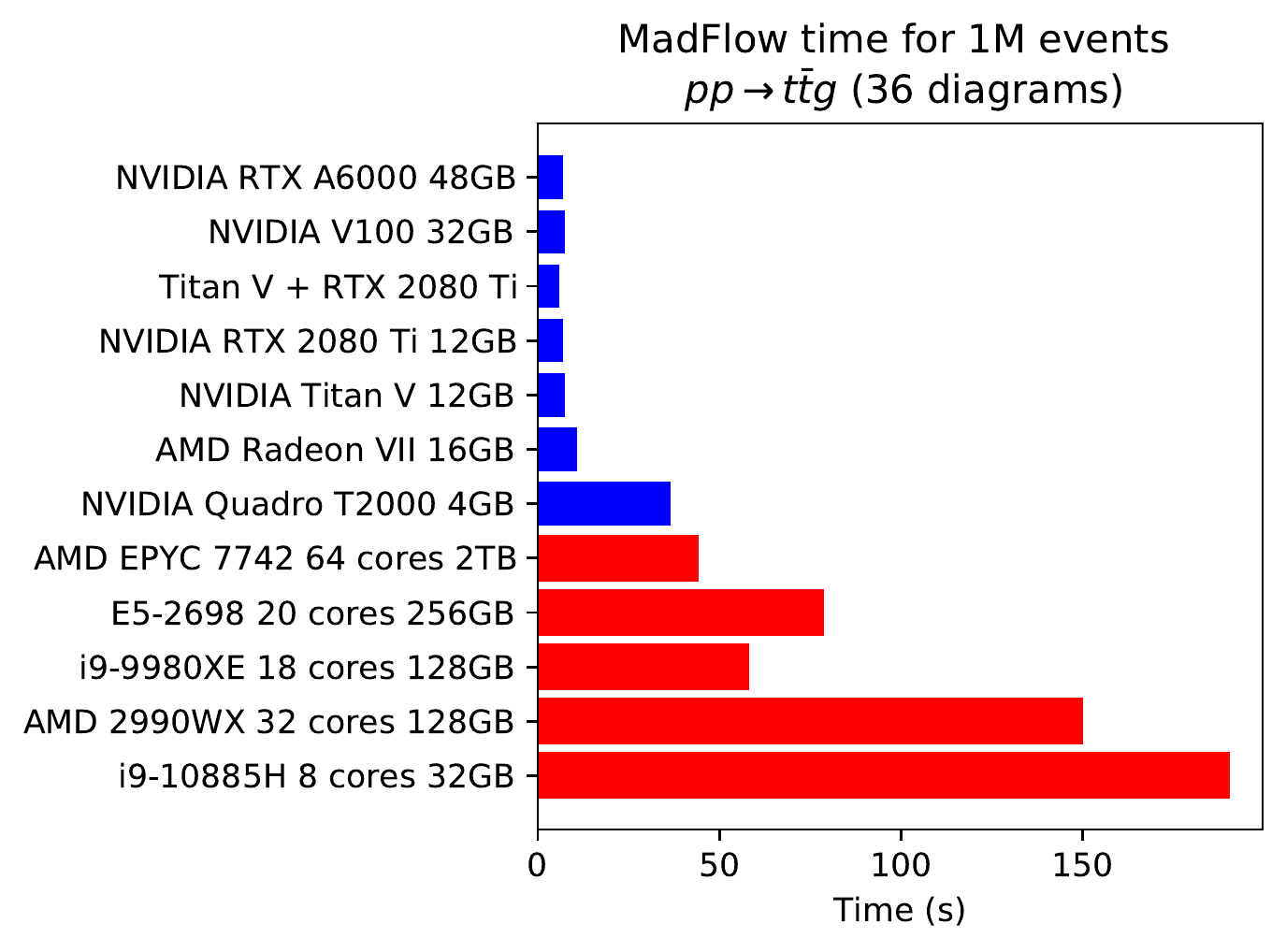}
    \hfill
    \includegraphics[width=0.49\textwidth]{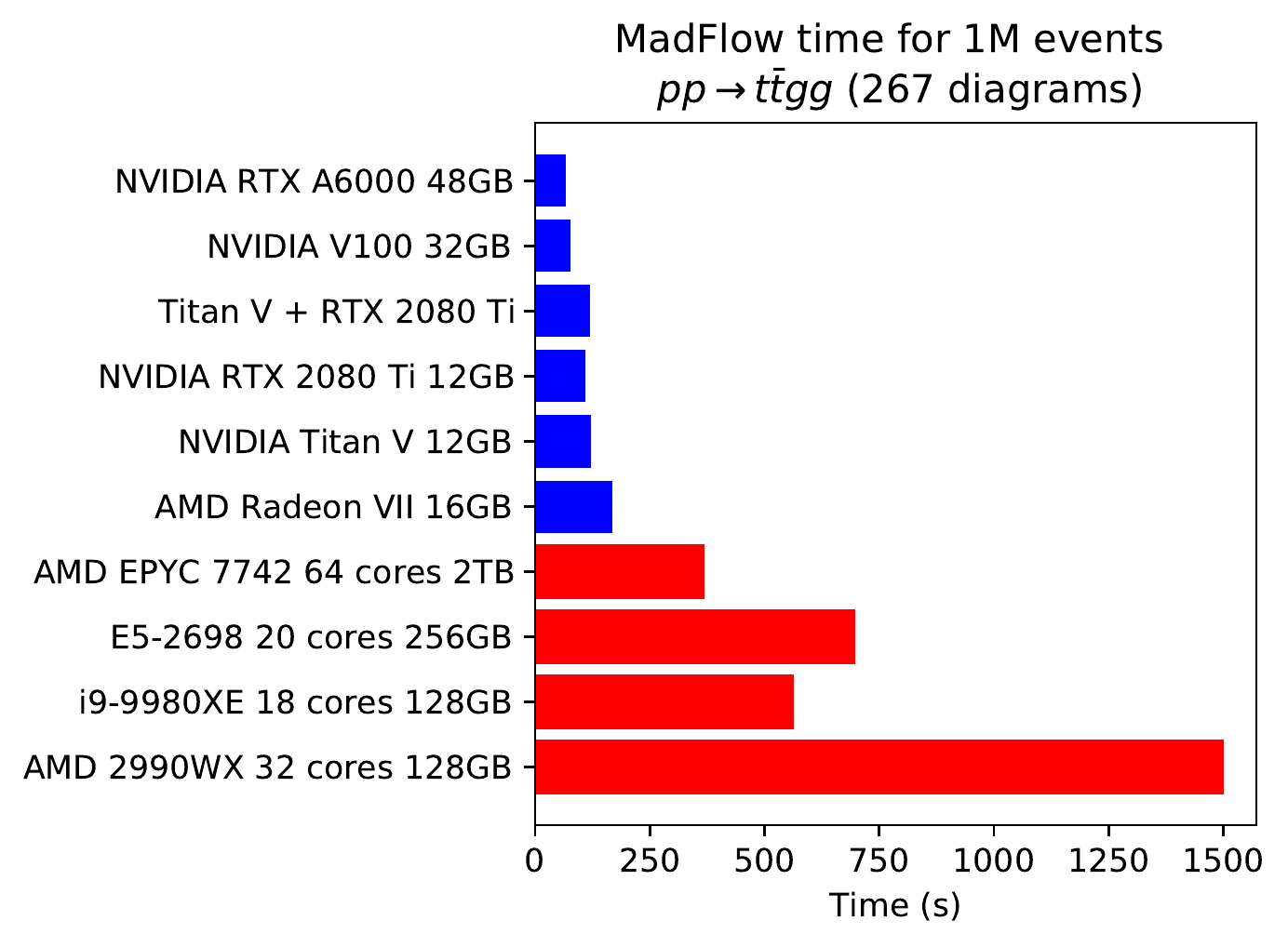}
    \caption{Benchmark of {\tt MadFlow} running on different devices. CPUs at the bottom (red) and GPUs at the top (blue). All the GPUs perform better than any of the CPUs for the processes being considered.}
    \label{fig:recall}
\end{figure}

The technical details of \texttt{MadFlow} have been discussed elsewhere~\cite{Carrazza:2021gpx}, the code is open source~\cite{juan_m_cruz_martinez_2021_4958257} and accessible at \href{https://github.com/N3PDF/madflow}{\color{blue} https://github.com/N3PDF/madflow}.
In Fig.~\ref{fig:recall} we recall some of the benchmarks performed in~\cite{Carrazza:2021gpx}.
As it can be appreciated, the improvements on performance are less pronounced as the number of diagrams grows (see Table~\ref{tab:oldbench}).

\begin{table}[ht]
    \centering
    \small
    \begin{tabular}{ l | c | c | c | c }
        Process & Diagrams & {\tt MadFlow} CPU & {\tt MadFlow} GPU & {\tt MG5\_aMC}  \\ \hline
        $pp \rightarrow t\bar{t}g$ & 36 & 57.84 $\mu$s & 7.54 $\mu$s & 93.23 $\mu$s \\
        $pp\rightarrow t\bar{t}gg$ & 267 & 559.67 $\mu$s & 121.05 $\mu$s & 793.92 $\mu$s \\
        \hline
    \end{tabular}
    \caption{Comparison of event computation time for {\tt MadFlow}
        and {\tt MG5\_aMC}, using an Intel i9-9980XE with 18 cores and 128 GB of RAM
        for CPU simulation and the NVIDIA Titan V 12 GB for GPU simulation. We observe a factor of $\sim$ 8 for the simpler calculation and only a factor of $\sim$ 4 for the more complicated one.}
    \label{tab:oldbench}
\end{table}

The results of Table~\ref{tab:oldbench} do not come as a surprise, a more complex computation leads to a more frequent context-switching and memory transference, both of which are very expensive operations in hardware accelerators.
This is in part driven by the growth of the memory footprint of the calculation.
The resulting effect is that fewer events can be computed in parallel leading to a loss of efficiency.
Since the performance of the calculation is connected to the number of diagrams, we will write for this exercise a low-level implementation of the matrix element using Cuda, 
due to the availability of GPUs manufactured by NVIDIA.

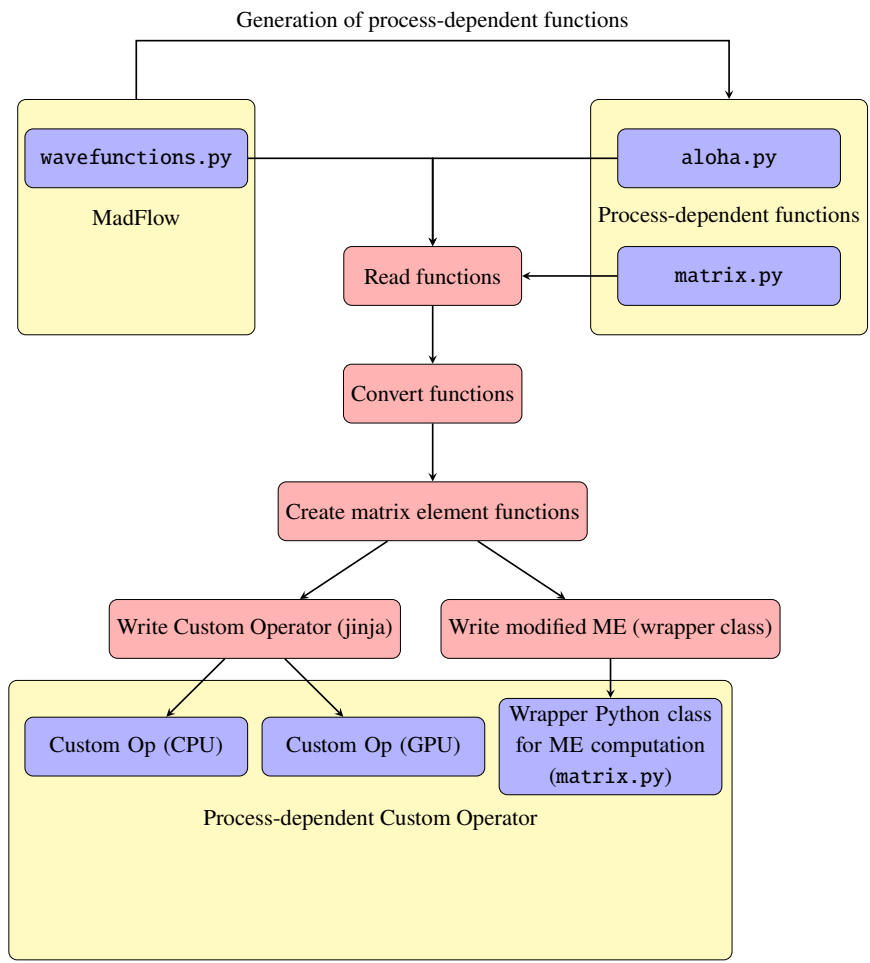
\begin{figure}[ht]
    \centering
    \scalebox{0.78}{
\tikzstyle{startstop} = [rectangle, rounded corners, minimum width=3cm, minimum height=1cm,text centered, draw=black, fill=red!30]
\tikzstyle{startstop2} = [rectangle, rounded corners, minimum width=3cm, minimum height=1cm,text centered, text width=6cm, draw=black, fill=red!30]
\tikzstyle{startstop3} = [rectangle, rounded corners, minimum width=3cm, minimum height=1cm,text centered, text width=3.5cm, draw=black, fill=red!30]
\tikzstyle{io} = [rectangle, rounded corners, minimum width=2cm, minimum height=1cm, text centered, text width=3.5cm, draw=black, fill=blue!30]
\tikzstyle{for} = [rectangle, minimum width=3cm, minimum height=1cm,text centered, draw=black, fill=orange!30]
\tikzstyle{if} = [diamond, minimum width=3cm, minimum height=1cm,text centered, draw=black, fill=green!30]
\tikzstyle{instance1} = [rectangle, rounded corners, minimum width=4cm, minimum height=4cm, text centered, draw=black, fill=yellow!30]
\tikzstyle{instance2} = [rectangle, rounded corners, minimum width=4.5cm, minimum height=4cm, text centered, draw=black, fill=yellow!30]
\tikzstyle{instance3} = [rectangle, rounded corners, minimum width=12.2cm, minimum height=4.75cm, text centered, draw=black, fill=yellow!30]
\tikzstyle{arrow} = [thick,->,>=stealth]
\def\wfx{0}
\def\wfy{0}
\def\alohax{10}
\def\matrixx{0}
\def\matrixy{-2}
\def\fileparsey{2}
\def\verticalspace{2}
\def\jinjaspacer{3}
\def\opspacer{2}
\def\xloop{4}

\begin{tikzpicture}

\node (source) [instance1] at ({\wfx},{\wfy - 1}) {MadFlow};
\node (product) [instance2] at ({\wfx + \alohax + \matrixx / 2},{\wfy - 1}) {Process-dependent functions};
\node (op_folder) [instance3] at ({\wfx + \alohax / 2 - 1.05},{\wfy - \fileparsey - 4*\verticalspace - 1.25}) {Process-dependent Custom Operator};


\node (wf) [io] at ({\wfx},{\wfy}) {\texttt{wavefunctions.py}};
\node (aloha) [io] at ({\wfx + \alohax},{\wfy}) {\texttt{aloha.py}};
\node (matrix) [io] at ({\wfx + \alohax + \matrixx},{\wfy + \matrixy}) {\texttt{matrix.py}};

\node (read_f) [startstop] at ({\wfx + \alohax / 2},{\wfy - \fileparsey}) {Read functions};

\node (parse_f) [startstop] at ({\wfx + \alohax / 2},{\wfy - \fileparsey - \verticalspace}) {Convert functions};

\node (create_me) [startstop] at ({\wfx + \alohax / 2},{\wfy - \fileparsey - 2*\verticalspace}) {Create matrix element functions};

\node (write_op) [startstop] at ({\wfx + \alohax / 2 - \jinjaspacer},{\wfy - \fileparsey - 3*\verticalspace}) {Write Custom Operator (jinja)};

\node (op_cpu) [io] at ({\wfx + \alohax / 2 - \opspacer - \jinjaspacer},{\wfy - \fileparsey - 4*\verticalspace}) {Custom Op (CPU)};
\node (op_gpu) [io] at ({\wfx + \alohax / 2 + \opspacer - \jinjaspacer},{\wfy - \fileparsey - 4*\verticalspace}) {Custom Op (GPU)};

\node (write_me) [startstop] at ({\wfx + \alohax / 2 + \jinjaspacer},{\wfy - \fileparsey - 3*\verticalspace}) {Write modified ME (wrapper class)};
\node (modified_me) [io] at ({\wfx + \alohax / 2 + \jinjaspacer},{\wfy - \fileparsey - 4*\verticalspace}) {Wrapper Python class for ME computation (\texttt{matrix.py})};

\draw [arrow] (source) -- ({\wfx},{\wfy + 2}) -- node[above] {Generation of process-dependent functions} ({\wfx + \alohax},{\wfy + 2}) -- (product);

\draw [arrow] (wf) -| (read_f);
\draw [arrow] (aloha) -| (read_f);

\draw [arrow] (read_f) -- (parse_f);

\draw [arrow] (parse_f) -- (create_me);
\draw [arrow] (matrix) -- (read_f);

\draw [arrow] (create_me) -- (write_op);
\draw [arrow] (write_op) -- (op_cpu);
\draw [arrow] (write_op) -- (op_gpu);

\draw [arrow] (create_me) -- (write_me);
\draw [arrow] (write_me) -- (modified_me);

\end{tikzpicture}}
    \caption{Architecture of the \texttt{MadFlow} framework, extended with the automatic generation of custom code written using the \texttt{jinja} template language with the {\tt MG5\_aMC} matrix element as input.}
    \label{fig:diagram}
\end{figure}

In order to maintain the number of changes to a minimum, the Cuda code is automatically transpiled from the \texttt{MadFlow} code (which is itself already parallelized on events).
A schematic diagram of the structure of the framework is shown in Fig.~\ref{fig:diagram}.
Only the necessary quantities (vertices, spinors or propagators) are generated as separated Cuda kernels.
The communication between the CPU and GPU kernels is done utilizing 
the C++ interface provided by TensorFlow from which one can launch Cuda kernels normally.

The modules that will launch the GPU-kernels are wrapped as TensorFlow custom operators.
Since we have not implemented the derivative of these operators, we have lost the possibility of taking the derivatives of the matrix element, however it would be trivial to extend the procedure to include that possibility.

\section{Benchmarks}
\label{sec:benchmarks}
We now compare the performance of the default \texttt{MadFlow}, without device-specific optimization, with the new version.
This option can be enabled by using the \texttt{--custom\_op} flag.

\begin{wrapfigure}{l}{0.42\textwidth}
    \centering
    \includegraphics[width=0.42\textwidth]{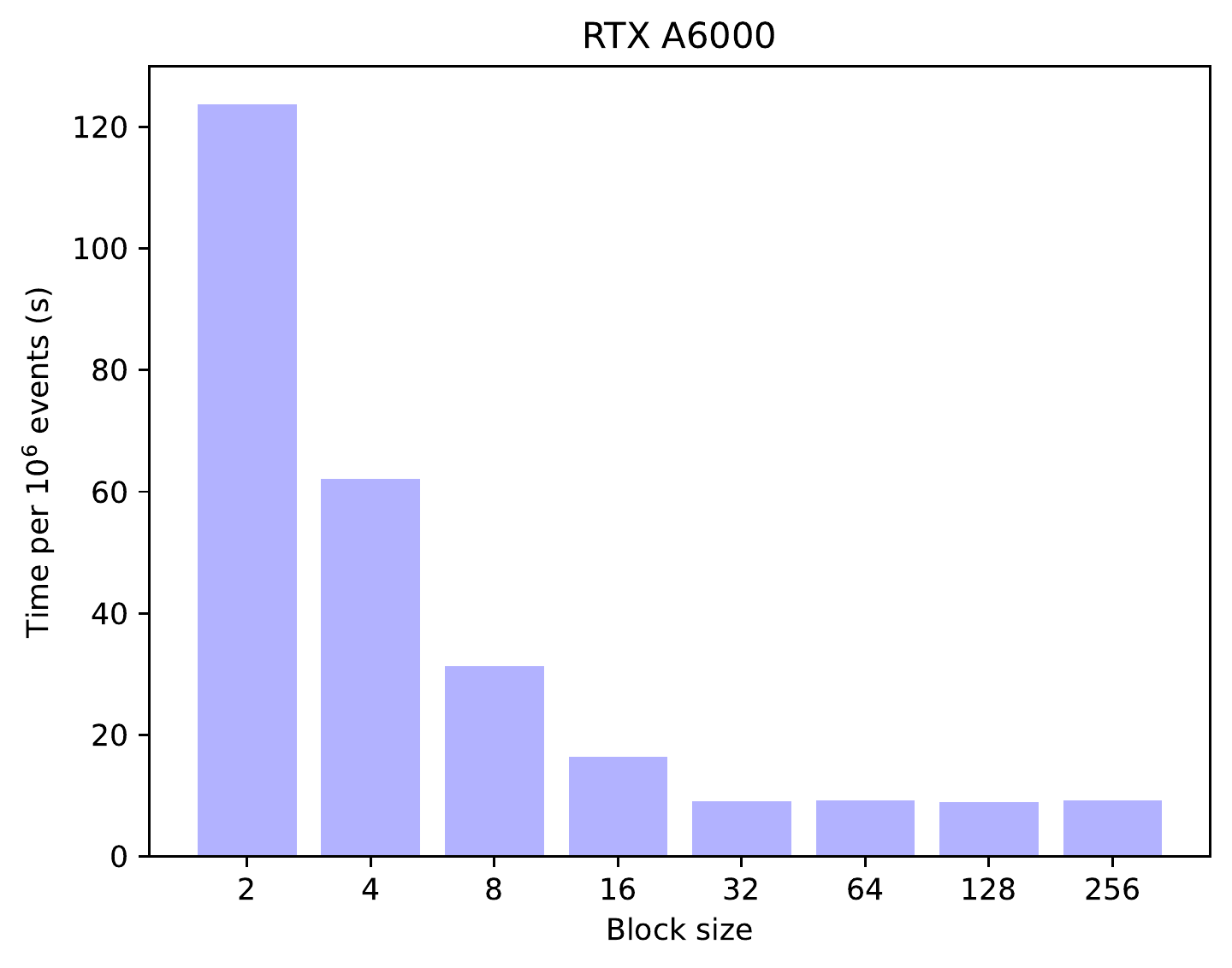}
    \caption{Time to run $10^{6}$ for a Leading Order $g g \rightarrow t \bar{t} g g$ calculation using \texttt{MadFlow} as a function of the number of threads in each block. We observe how a bad choice of the block size can have a huge impact on the performance of the calculation.}
    \label{fig:blocksthreads}
\end{wrapfigure}

An example on how a low-level interface gives the user a finer control of the calculation can be seen in Fig.~\ref{fig:blocksthreads}.
The computing architecture of NVIDIA GPUs divides the computation in blocks, each of these blocks is then further divided into threads.
How the computation is spread on these computing units has a huge effect on the efficiency of the calculation.

By having direct access to the underlying hardware one might be able to outperform the more generic settings deduced automatically by a higher-level language.
For instance, two different graphic cards of the same manufacturer might have very different characteristics and so a one-size fits-all approach might not take full advantage of the hardware.
As seen in Fig.~\ref{fig:blocksthreads}, there is a range of block-sizes that achieve similar performance, but this might depend on the ability of the GPU (or the underlying framework) to hide latency by for instance transferring memory while the previous block computation is still ongoing.

\begin{figure}[b]
    \includegraphics[width=0.49\textwidth]{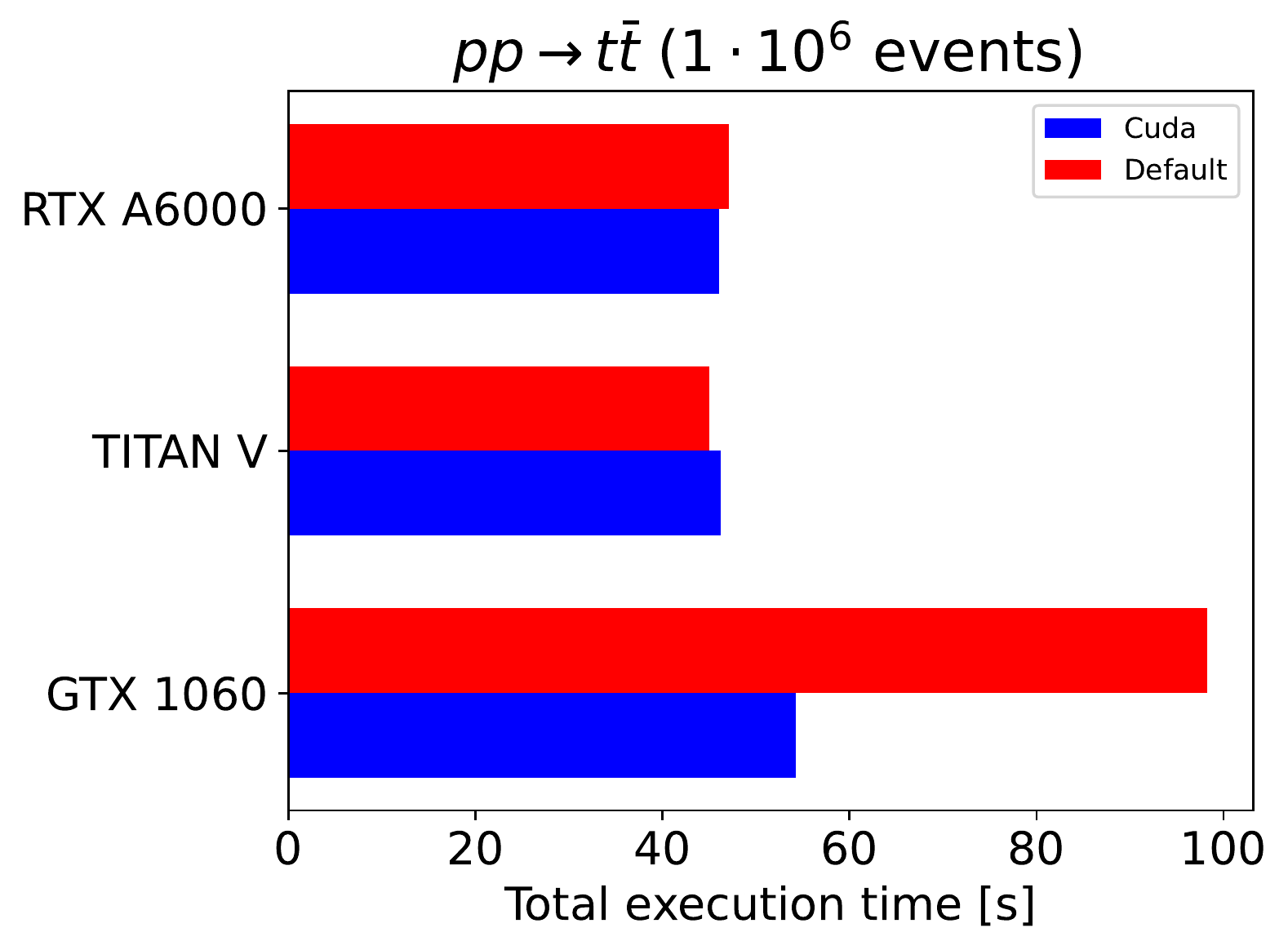}
    \hfill
    \includegraphics[width=0.49\textwidth]{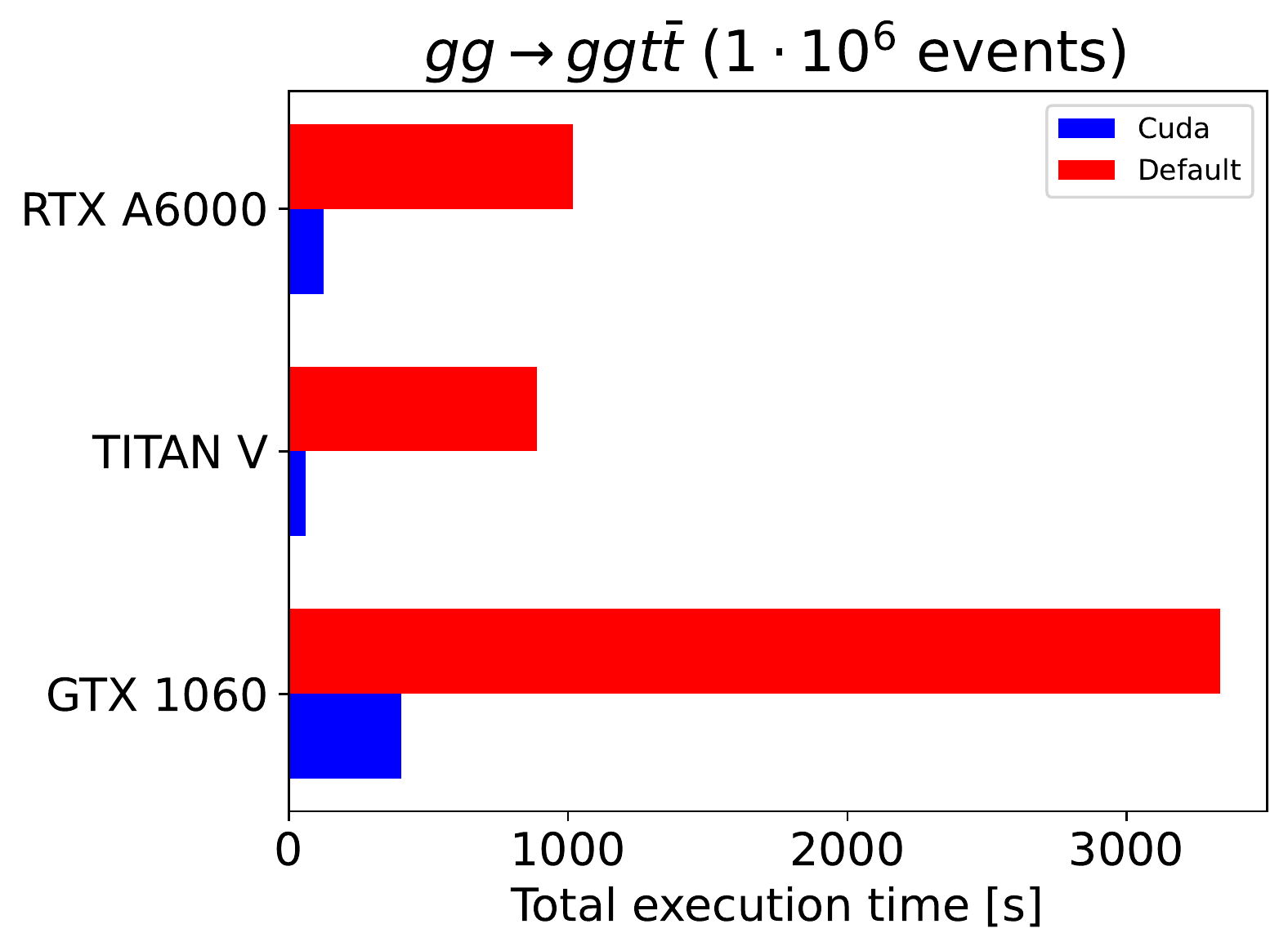}
    \caption{Benchmark of the default \texttt{MadFlow} (red) against the Cuda-extended version (blue) for the matrix element. In the left plot we observe no difference between the default autogenerated code and the custom operator written in Cuda for the better GPUs, while we find some difference for a low-end GPU. In the right plot instead we add two extra gluons to the process, increasing the number of diagrams being computed. In this case we find a much bigger difference for all the three GPUs tested.}
    \label{fig:finalbench}
\end{figure}

In Fig.~\ref{fig:finalbench} we compare the time it takes to compute two different processes on different hardware with and without custom code.
We observe that in the simplest of the two calculations ($pp \rightarrow t\bar{t}$) there is only a performance gain when using a lower end GPU, while for the high-end GPUs there is no appreciable loss of efficiency by using a more generic approach.
Being able to have a finer control of the computation allow us to overcome some limitations of a lesser hardware.
When the computation becomes more complicated we find that the generic approach becomes worse even for the high-end GPUs.
It is interesting to note that they still perform much better than a low-end GPU, and, as a consequence, the gains of the custom code are also more important in the low-end case.

In conclusion, the choice of sacrificing the convenience of hardware-agnostic code depends on both the type of hardware available and the specific calculation at hand.
The tools presented in this note aim to be flexible enough that both approaches are possible.

\acknowledgments
SC and JCM are supported by the European Research Council under the European
Union's Horizon 2020 research and innovation Programme (grant agreement number
740006).

\scriptsize
\bibliographystyle{../JHEP}
\bibliography{../blbl}
\end{document}